\documentclass[aps,showpacs,twocolumn,prl,superscriptaddress]{revtex4-1}
\usepackage{graphicx,color,epsfig,rotate}
\usepackage{color}
\usepackage{graphicx}% Include figure files
\usepackage{dcolumn}% Align table columns on decimal point
\usepackage{amssymb}
\usepackage{amsmath}
\usepackage{amsfonts}
\usepackage{docs}%
\usepackage{bm}%
\usepackage[colorlinks=true,linkcolor=blue,pagecolor=blue,filecolor=blue,menucolor=blue,urlcolor=blue,citecolor=blue,anchorcolor=blue]{hyperref}%
\usepackage{graphics}
\usepackage{standalone}
\usepackage[toc,page]{appendix}
\usepackage{import}
\usepackage{enumerate}
\usepackage{sidecap}

\newcommand{\bsigma}{{\boldsymbol \sigma}}

\newcommand{\btau}{{\boldsymbol \tau}}
\begin{document}

% ==============================================================================

\title{Superconductivity in Type-II Weyl Semimetals}

\author{M. Alidoust}
\affiliation{Department of Physics, K.N. Toosi University of Technology, Tehran 15875-4416, Iran}

\author{K. Halterman}
\affiliation{Michelson Lab, Physics Division, Naval Air Warfare Center, China Lake, California 93555}

 \author{A. A. Zyuzin}
   \affiliation{Department of Physics, KTH-Royal Institute of Technology, Stockholm, SE-10691 Sweden}
\affiliation{Ioffe Physical--Technical Institute,~194021 St.~Petersburg, Russia}

% ==============================================================================

%\date{\today}

\begin{abstract}
We study superconductivity in a Weyl semimetal with a tilted dispersion around two Weyl points of opposite chirality. In the absence 
of any tilt, the state with zero momentum pairing between two Fermi sheets enclosing each Weyl point has four point nodes in the superconducting gap function. Moreover, the surface of the superconductor hosts Fermi arc states and Majorana flat bands. 
We show that a quantum phase transition occurs at a critical value of the tilt, at which two gap nodes disappear by merging at the center 
of the first Brillouin zone,
or by escaping at its edges, depending on the direction of the
tilt. The region in the momentum space that the 
 Majorana flat band
occupies is found to 
increase as the tilt parameter is made larger. 
Additionally, the superconducting critical temperature and electronic
specific heat can be 
enhanced in the vicinity of the quantum phase transition due to 
the singularity in the electronic 
density of states.
\end{abstract}
\pacs{ 74.25.Dw, 03.65.Vf, 71.90.+q} 
%74.25.Dw	Superconductivity phase diagrams
%03.65.Vf		Phases: geometric; dynamic or topological
%71.90.+q	Other topics in electronic structure (restricted to new topics in section 71)

\maketitle

% ==============================================================================

% ==============================================================================

\section{introduction}
Weyl semimetals and 
the
closely related Dirac semimetals are topologically nontrivial phases of matter
which were first predicted theoretically~\cite{Murakami, Weyl, Weyl1, Weyl2, Weyl3} and recently realized
experimentally in $\mathrm{TaP}$, $\mathrm{NbP}$, $\mathrm{TaAs}$, $\mathrm{NbAs}$, $\textrm{Cd}_3\textrm{As}_2$, $\mathrm{Na_3Bi}$, 
and $\mathrm{ ZrTe_5}$
~\cite{bib:WSM2, Xu294, PhysRevX.5.031013, bib:WSM1, CdAs, Swiss_Weyl, Xu300, Xu613, Felser, Kharzeev, Xiong413}. 
The band structure of these materials contains nondegenerate 
conduction and valence bands touching at certain
``Weyl points", 
around which, the energy dispersion of the
quasiparticle excitations is linear. The corresponding three
dimensional conical band structure in this part of the spectrum is
referred to as a Weyl cone. 
The Weyl  points which always come in pairs, represent a source or a
sink of Berry curvature in the momentum space, and describe Weyl fermions in real space~\cite{Volovik}. Although the band structure in the bulk of Weyl semimetals is gapless, the bulk-edge correspondence gives rise to Fermi arc surface states
that connect pairs of Weyl points. %kh

It was found that the tilt of the Weyl cone results in %the
a Lifshitz transition,
whereby the point-like Fermi surface transforms into
a series of
electron and hole Fermi pockets connected to each other by a single point~\cite{vol14}. 
This is %now
the so-called type-II phase of a Weyl semimetal~\cite{sol15}. 
The tilt of the Weyl cone can be realized, for example, by applying strain to the crystalline Weyl semimetal structure~\cite{bib:WSM_HgTe} or 
a Zeeman field ~\cite{Weyl2_Zeeman}. The electronic transport in Weyl semimetals with tilted Weyl cones including conductance \cite{PhysRevB.91.115135}, chiral anomaly~\cite{sol15}, and intrinsic anomalous Hall effect~\cite{ZyuzinRakesh} become anisotropic compared 
to an ideal Weyl semimetal. 

Superconductivity in Weyl metals may give rise to topological phases that host Majorana fermions.
This has
triggered a great deal of interest both experimentally and theoretically 
due to the possible applications in quantum computation \cite{Alicea}.
In general, one expects zero-momentum inter-valley pairing and  finite momentum intra-valley pairing in Weyl metals
due to the 
multiple number of valleys within its band structure~\cite{PhysRevB.86.054504, PhysRevB.86.214514, Tanaka, Weyl_SC_ours, bib:Haldane_SC}. 
It was argued that inter-valley pairing 
can be energetically more favorable than
intra-valley pairing,
provided 
time reversal symmetry is violated, but inversion symmetry is preserved~\cite{Weyl_SC_ours}. 
If such a condition is realized, the superconducting gap exhibits four nodes with Majorana surface states that connect projections of the node locations onto the surface of the 
Brillouin zone, as well as  Fermi arc surface states 
that connect two distinct Fermi sheets~\cite{PhysRevB.86.054504, Tanaka, Weyl_SC_ours, bib:Haldane_SC}.

In this paper, 
our objective is to theoretically investigate superconductivity in type-II Weyl metals. 
Motivated primarily 
by the recent experimental observations
of a type-II Weyl semimetal state~\cite{Type2Exp_New1, Type2Exp_New2, Type2Exp_New3, Type2Exp_New4, WT2_16, Efremov},
and pressure enhanced superconductivity~\cite{Weyl_SC_exp} in the
type-II Weyl semimetal candidate $\mathrm{MoTe}_2$, we 
investigate the zero momentum inter-valley superconducting state in 
a Weyl semimetal with tilted conical spectrum around 
the Weyl points. 
We show that if the cone tilts in 
same direction as along the two Weyl points, then 
there can be a critical tilt orientation in which a quantum phase transition  
occurs.  
At this phase transition, the two nodes in the
superconducting gap function disappear
by merging at the center 
of the first Brillouin zone 
or by escaping at the Brillouin zone edges,
depending on whether the
Weyl cones are tilted towards or away from 
each other. Interestingly, while the Fermi arc surface states tend to defined
hybridize with the bulk bands in the Weyl superconductor, the region  
of momentum space where the Majorana band 
is flat becomes expanded by increasing
the tilt angle.
The phase transition can 
also have a number 
of profound effects, including
an enhancement of 
 the superconducting critical temperature and electronic specific heat arising from the Van-Hove singularity in the electronic 
density of states.

The rest of the paper is organized as follows: In Sec.~\ref{sec:First}, we first introduce the low-energy model
used for a 
Weyl superconductor with  tilted conical spectrum, 
and calculate the temperature at which the superconductor-metal phase transition occurs for various limits of the tilt. In Sec.~\ref{sec:Second} we analyze the spectrum of both the bulk and surface states in this system analytically.
In Sec.~\ref{sec:theor}, we extend our analysis of the surface states by implementing 
exact numerical and diagonalization techniques. 
Finally, in Sec.~\ref{sec:conc}, we present a summary of our results and give  concluding remarks. 

\section{Model and Thermodynamic Properties}\label{sec:First}
We consider a minimal model for a Weyl semimetal with only two valleys in the band structure and
 a Weyl point in each valley~\cite{Weyl2}.  %kh
This model can be described by $H = \int \frac{d^3k}{(2\pi)^3}c^{\dag}_{\mathbf{k}}H(\mathbf{k}) c_{\mathbf{k}}$, where
the momentum space Hamiltonian is given by,
\begin{equation}
H(\mathbf{k}) = v_z(\sigma_z + \alpha\tau_z+\beta)(k_z \tau_z- Q) + v_{\perp}[\hat{\bm z}\times \bsigma] \cdot \mathbf{k},
\label{eq:Hamkz}
\end{equation}
where $\bsigma$  and $\btau$
are vectors composed of the three Pauli matrices acting on 
the spin and valley degrees of freedom respectively, and
$2Q$ is the separation of the two Weyl cones along the $z$-axis in the
momentum space.
The parameters $\alpha$ and $\beta$ describe the tilt of the two Weyl cones 
that are either in the same direction or the
opposite direction with respect to each other,  and
$v_z, v_{\perp}$ are the components of the anisotropic Fermi velocity 
when the tilt parameters vanish.
We represent the annihilation operator $c_{\mathbf{k}}$ as $c_{\mathbf{k}}\equiv
[c_{\mathbf{k},\uparrow,R},c_{\mathbf{k},\downarrow,R},c_{\mathbf{k},\uparrow,L},c_{\mathbf{k},\downarrow,L}]^{\mathrm{T}}$, 
for electrons with momentum $\mathbf{k}$ and spin $\uparrow, \downarrow$, in 
the two valleys denoted by $R, L$. 

We consider the case where the direction of tilt is along the
direction in which the two Weyl points are separated, resulting  in  nontrivial  behavior of the nodes in the gap function of 
the Weyl superconductor. We also consider the limit $\alpha=0$, focusing on the effects of tilting the Weyl cones in opposite directions with respect to each other. The effects of finite values of $\alpha$ are discussed in Sec.~\ref{sec:conc}.

Let us
recall the band structure of a Weyl semimetal as a function of the tilt parameter. For slight tilting 
with $|\beta|<1$, the band-structure of the system consists of two valleys 
where the electron and hole bands touch each other at a Weyl point in both valleys. These two point nodes are separated by $2Q$ along 
the $z$-axis in the momentum space. With increasing tilt, one can reach a situation where $|\beta|=1$, at which 
the two Weyl points transform into a nodal line that coincides with  the $z$-axis. 
By increasing the tilt further such that $|\beta|>1$, 
the nodal line transform into electron and hole Fermi pockets connected to each other by the point nodes.

We now consider inversion symmetric inter-valley s-wave superconducting pairing, which can be described by $H = \frac{1}{2}\int \frac{d^3k}{(2\pi)^3}\psi^{\dag}_{\mathbf{k}}\mathcal{H}(\mathbf{k}) \psi_{\mathbf{k}}$, where 
$\psi_{\mathbf{k}}=[c_{\mathbf{k}},c^{\dag}_{\mathbf{-k}}]^{\mathrm{T}}$ is the Nambu spinor and the BCS Hamiltonian is given by
\begin{eqnarray}\label{Main_H}
\mathcal{H}(\mathbf{k}) =\left( \begin{array}{cc}
H(\mathbf{k})-\mu& i\sigma_y\tau_x\Delta \\
 -i \sigma_y\tau_x\Delta^{\ast}& -H^{\mathrm{T}}(-\mathbf{k})+\mu
\end{array} \right),
\end{eqnarray}
where  $\mu$ is the Fermi energy and $\Delta$ 
is the spatially homogeneous  inter-valley $s$-wave order parameter, which 
is defined by a solution of the self-consistency equation,
\begin{eqnarray}\nonumber\label{SCE}
\Delta &=& \frac{\lambda T}{4}\sum_{n\in \mathbb{Z}}\sum_{s=\pm1}\int\frac{d\mathbf{k}}{(2\pi)^3}\frac{\Delta}{\omega_n^2 + E_s^2(\mathbf{k})}\\
&\times&\bigg[1+ s\frac{v_z^2k_z^2}{\sqrt{|\Delta|^2 v_z^2 k_z^2+(\mu-\beta v_z k_z)^2\epsilon^2(\mathbf{k})}}\bigg].
\end{eqnarray}
Here $\lambda>0$ is the interaction constant, which is assumed to be independent from the tilt parameters, $T$ is the temperature, 
and $\omega_n=\pi T(2n+1)$ is the Matsubara frequency.
We also have $\epsilon(\mathbf{k}) = \sqrt{v_z^2k_z^2 + v_{\perp}^2(k_x^2+k_y^2)}$,
and dispersion of quasiparticles
\begin{eqnarray}\label{EV}
E_{s}^2(\mathbf{k})&=&%\alpha (k_z-Q) 
\epsilon^2(\mathbf{k})+|\Delta|^2+(\mu-\beta v_z k_z)^2\\\nonumber
&+&2s\sqrt{\epsilon^2(\mathbf{k})(\mu-\beta v_z k_z)^2 +v_z^2k_z^2|\Delta|^2},
\end{eqnarray}
which clearly has nodes along $z$-axis in momentum space at $\beta=0$. 

We emphasize that we are interested in the most simplest case of s-wave phonon mediated superconductivity in Weyl metals discussed in Ref. \cite{Weyl_SC_ours}.  Particularly, it was shown that the intra-valley pairing is less energetically favorable than the inter-valley one provided that the inversion symmetry is preserved.  
We now extend that theory to Weyl metals with tilted cones. The inversion symmetry is preserved if two Weyl cones are not tilted in the same directions with respect to each other, which allows us to conclude that the inter-valley pairing is the dominant pairing in the presence of tilt parameter $\beta$. 

\begin{figure}[t!]
\includegraphics[width=8.0cm]{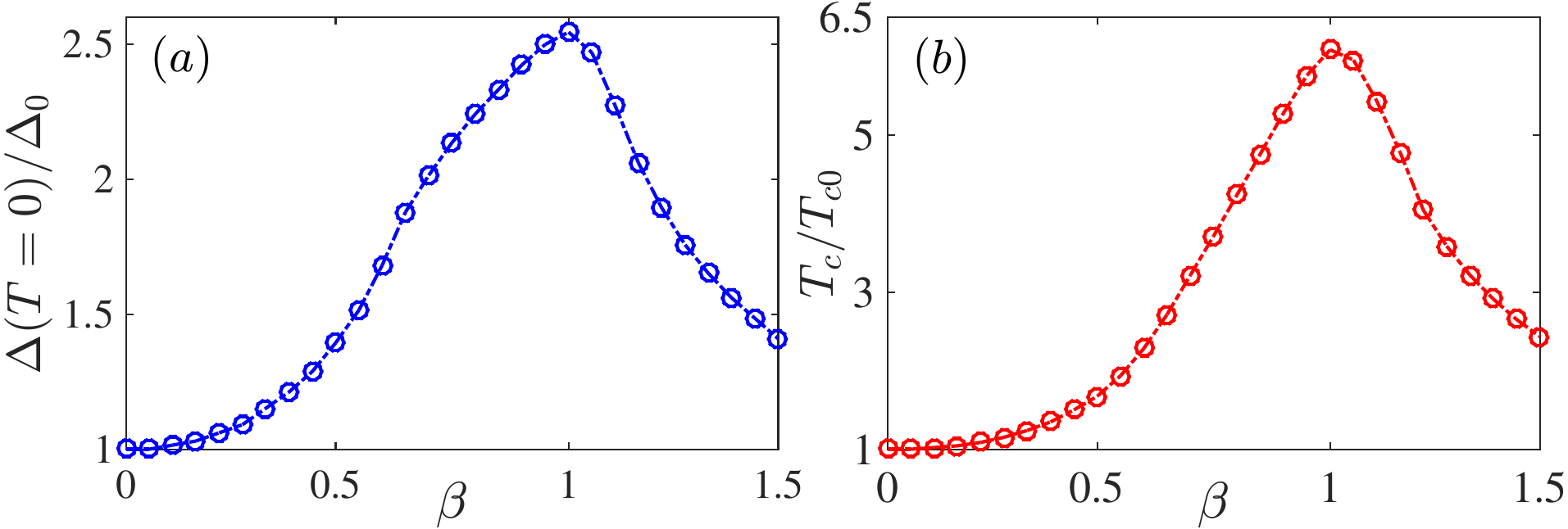}
\caption{(Color online) Tilt parameter $\beta$ dependence of: (a) order parameter $|\Delta|/|\Delta_0|$ at zero temperature and (b) critical temperature $T_c/T_{c0}$, which are normalized to their values in the case $\beta=0$. 
  }
\label{fig0}
\end{figure}

\subsubsection{Critical Temperature}
The density of states at the Fermi level is a function of the tilt parameter, which modifies the critical temperature and thermodynamic properties of the superconductor.
In particular, in the high temperature regime $T\gg |\Delta|$,
 the critical temperature can 
be increased by increasing the tilt parameter (provided $|\beta|\ll 1$). Solving Eq. \ref{SCE} in these limits, we find that
\begin{align}
T_c = \frac{2\omega_D \gamma}{\pi}\exp{[-6(1-6\beta^2/5)/\lambda\nu]},
\end{align}
where $\nu = \mu^2/ (2\pi^2 v_z v_{\perp}^2)$ is the electronic density of states at the Fermi energy 
per pseudospin and per valley in the absence of tilt and $\omega_D$ is the Debye frequency, which we assume to be independent 
from the tilt parameter. 
At large tilt $|\beta| \gg1$, the critical temperature exponentially 
vanishes with further increases of the tilt: 
\begin{align}
T_c = \frac{2\omega_D \gamma}{\pi}\exp{[-|\beta|/\lambda\tilde{\nu}]},
\end{align}
where $\tilde{\nu} = \frac{k_{0}^2}{8\pi^2 v_z}$ and $k_0$ is the momentum cut-off of the unbounded electron-hole pockets, which itself depends on the tilt parameters, \cite{sol15}. Such a cut-off is an artifact of the linearized model only, and thus does not arise in our tight-binding calculations given is Section \ref{sec:theor}.  Dependence of the critical temperature $T_c$ and the gap function $|\Delta|$ on the tilt parameter $\beta$ is shown in Fig. \ref{fig0}.

We also note that the important corrections to $T_c$ might come from the many body effects in the case of large tilts. Indeed, the tilt increases the density of states at the Femi level. Thus, the tilt modifies both the electron-phonon interaction parameter and the screening of the electron-electron interaction potential. 
We consider vertex corrections for the electron-phonon scattering within Migdal approximation $\omega_D/\mu \ll1$. However, the increase of the density of states increases the screening of the electron interaction potential and thus suppresses the critical temperature in the spirit of McMillan theory. Atomistic and ab initio computations may help to understand more about these complicated issues. We believe that a detailed answer to this fundamental question with accurate treatment of these two interactions goes beyond the scope of our paper. 

\subsubsection{Specific Heat}
One of the most fundamental physical quantities in superconductivity is the electronic specific heat,
$C$, which at low temperatures can be written as
\begin{equation}
C(T) = 2V\sum_{s=\pm 1}\int \frac{d^3k}{(2\pi)^3} \left[\frac{E_s(\mathbf{k})/2T}{\cosh (E_s(\mathbf{k})/2T)}\right]^2,
\end{equation}
where $V$ is the volume of sample.
In this regime, the specific heat is proportional to the third power of the temperature,
$
C(T) = \frac{7\pi^2}{15} \frac{\nu V}{|\Delta|^2} T^3(1+6 \beta^2),
$
reflecting the well known behavior for superconductors with point nodes in the gap \cite{Mineev}. If  the Weyl cones are strongly tilted ($|\beta|\gg 1$), and we again consider the low
temperature regime ($T\rightarrow 0$), this leads to a fully 
gapped superconducting state, such that the specific heat is now exponentially suppressed 
and  described by the asymptotic behavior:
$
C(T)\propto \tilde{\nu} |\Delta|^{5/2}\frac{e^{-|\Delta|/T}}{|\beta| T^{3/2}}.
$

\begin{figure*}[t!]
\includegraphics[width=18.0cm,height=5.20cm]{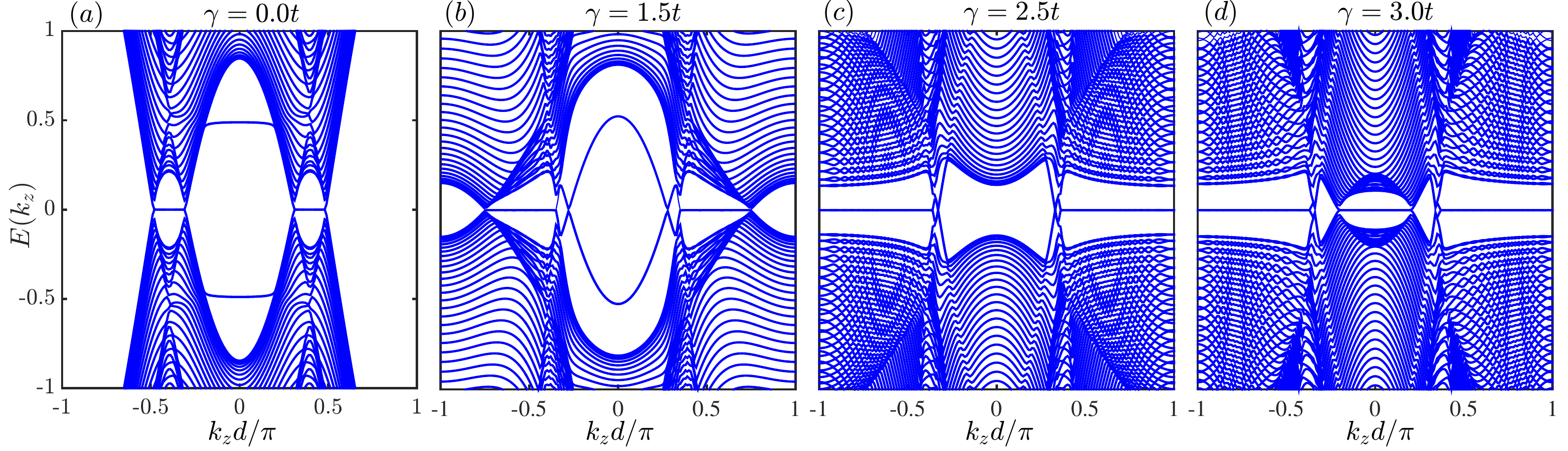}
\caption{(Color online) Band structure of the bulk Weyl
  superconductor as a function of momentum $k_z$ for different values of the tilt parameter $\gamma$
  (see Eq.~(\ref{eq1})). 
  The energy is normalized by the hopping energy integral $t$. 
  The 
  $x$-axis is directed normal to
  the  two parallel plane boundaries of the superconductor, so that
  the longitudinal momenta $k_y$ (set to zero here),
  and $k_z$ are good quantum numbers. 
  %Here, to obtain the band structure, we set $k_y=0.0$ for simplifying our analysis. 
  The left panel shows the band structure of a Weyl superconductor without tilting, where $\gamma=0$. The $\gamma$
  parameter is increased from left to right as follows: $\gamma=0, 1.5t, 2.5t$, and $3t$. The chemical potential is fixed at $\mu=0.5t$ and
  the superconducting gap is set at $\Delta=0.2t$. As seen, the region of the Majorana zero energy flat band, bounded to the nodes in the superconducting gap, increases as $\gamma$ increases. 
  When $\gamma =2.0t$ there is a quantum phase transition, whereby two out of the four gap nodes migrate to 
  the edges of the first Brillouin zone and then reappear at the center when $\gamma\approx 2.7t$. 
  With the further increase of $\gamma$, the Majorana flat band region increases.
  }
\label{fig2}
\end{figure*}

\section{Bulk and surface states}\label{sec:Second}
In this section we analyze the quasiparticle spectrum in the bulk of a
Weyl superconductor and discuss the presence of any nontrivial surface states. 
Using the expression in Eq.~(\ref{EV}), the eigenvalues of the Hamiltonian in Eq.~(\ref{Main_H}) can be written as 
$%\begin{equation}\label{dispersion_WSC}
\mathcal{E}_s(\mathbf{k})=E_{s}(\mathbf{k}\mp Q \mathbf{e}_z)
$. %\end{equation}
Without loss of generality, we  consider the part of the spectrum $E_{s}(\mathbf{k} - Q \mathbf{e}_z)$, which describes the
excitations of quasiparticles in the vicinity of the Fermi sheet enclosing the Weyl point at $\mathbf{k}=(0,0,Q)$. A similar approach 
can be applied
to the Fermi sheet enclosing the Weyl point at $\mathbf{k}=(0,0,-Q)$.

We first examine the limit $\mu=0$. 
For $|\beta|<1$, the spectrum of 
quasiparticles in the 
superconductor around the
point  $\mathbf{k}=(0,0,Q)$ 
has two point nodes in the gap function 
at %kh
$\mathbf{k}=(0,0,k_{z,\pm})$, where 
\begin{equation}
k_{z,\pm} =Q \pm \frac{|\Delta|}{v_z\sqrt{1-\beta^2}}.
\end{equation} 
%We note that superconducting order parameter $\Delta$ is itself a function of $\beta$, as follows from Eq. \ref{SCE}.
The increase of the tilt parameter %in the limit 
for $|\beta|<1$ causes $|\Delta|$ to increase as well as the distance between the nodes. The spectrum along the direction of the tilt is given by $E=\pm v_z(1-\beta^2)\delta k_{z,\pm}$, where $\delta k_{z,\pm} =k_z- k_{z,\pm}$.
When the Weyl cone is over-tilted, such that $|\beta|\geq 1$, the spectrum of quasiparticles in the superconductor is gapped. 
%We note that $\Delta$ is a function of tilt parameter $\beta$.

We now consider the case $\mu>0$. In the 
range $|\beta|< 1$, the  two point nodes in the energy spectrum 
are located at $\mathbf{k}=(0,0,k_{z,\pm})$, where now
\begin{equation}
k_{z,\pm} =Q +\frac{\mu}{v_z(1-\beta^2)}\left[\pm \sqrt{1 + (1-\beta^2)\frac{|\Delta|^2}{\mu^2}}-\beta \right].
\end{equation}
The two nodes in the gap function are separated in the momentum space by $|k_{z,+}-k_{z,-}|$. 
With the increase of the tilt parameter $|\beta|$, this separation grows. 
For positive (negative) $\beta$, the 
position of the node at $k_{z,+}$ ($k_{z,-}$) gets closer to the Weyl point $\mathbf{k}=(0,0, Q)$, while the position of the node at $k_{z,-}$ $(k_{z,+})$ diverges to $k_{z,-}\rightarrow -\infty$ ($k_{z,+}\rightarrow \infty$) as $|\beta| \rightarrow 1$. At $|\beta| = 1$, 
there is only one node at,
\begin{equation}
k_{z}=Q+ \beta\bigg(1+\frac{|\Delta|^2}{\mu^2}\bigg) \frac{\mu}{2v_z} ,
\end{equation}
at which the spectrum of quasiparticles in the tilt direction in the linear in momentum approximation is given by $\pm  v_z\frac{2\mu^2 \delta k_z}{\mu^2+|\Delta|^2}\Theta(\mu^2-|\Delta|^2)$. Here we note that $\Delta$ is a function of the tilt parameter.
The solution for the second node diverges, which means that the linearized model of the Weyl semimetal is no longer
 applicable in this regime. Instead, one has to take
into account the proper band structure of the entire first Brillouin zone so that this spurious divergence can 
be cut by the momentum corresponding to the edge or center  
of the first Brillouin zone. We investigate this issue numerically in the next section. 

Theoretically, a second node reappears 
by further increasing the tilt parameter in the region 
$\sqrt{1+\mu^2/|\Delta(\beta)|^2}>|\beta|>1$, 
such that the distance between the nodes $|k_{z,+}-k_{z,-}|$
decreases. 
Finally, the two nodes merge together when the tilt parameter has the value,
$
|\beta|=\sqrt{1+\mu^2/|\Delta(\beta)|^2},
$
after which a gap opens up in the energy spectrum. We believe that last two conditions could be realized
due to proximity effect in the superconductor- Weyl semimetal junctions.

We now briefly discuss the surface states of the Weyl superconductor.
There %kh There or These?
are Majorana zero energy edge modes, which exist along %kh on
 an arc, connecting the
nodes of the gap function at $\mathbf{k}=(0, 0, k_{z,-})$ and $\mathbf{k}=(0, 0, k_{z,+})$. 
For example, at the surface perpendicular to the $x$-direction, the energy spectrum of the Majorana edge modes 
as a function of the longitudinal momenta $k_y$ and $k_z$,  
 is given by,
\begin{equation}\label{MFS}
E_{M}=\frac{|\Delta| v_z (k_z-Q) v_{\perp} k_y}{(\mu-\beta v_z(k_z-Q))\sqrt{(\mu-\beta v_z(k_z-Q))^2-v_{\perp}^2 k_y^2}},
\end{equation}
which is evaluated away from the gap nodes at $\mathbf{k}=(0,0,k_{z,\pm})$,
and satisfies the conditions  $v_{\perp}^2k_y^2< (\mu-\beta v_z(k_z-Q))^2$, and $|\Delta|<|\mu-\beta v_z(k_z-Q)|$.
In the limit $\beta =0$, this expression coincides with the one found
in Ref.~\onlinecite{Tanaka}. 
There are two Majorana zero energy flat bands defined by $k_y=0$ and $k_z=Q$. Tilting of the Weyl cones increases the 
regions of flat bands in the momentum space along the $z$-axis as $|k_{z,+}-k_{z,-}|$, while the region 
of  flat bands along the $y$-axis in the momentum space diminishes.
Having analyzed a few relevant and simple cases,
we are now in a position 
to consider more complicated effects by
studying  the spectrum of the Majorana flat bands and Fermi arcs numerically.

\begin{figure*}[t!]
\includegraphics[width=18.0cm,height=5.30cm]{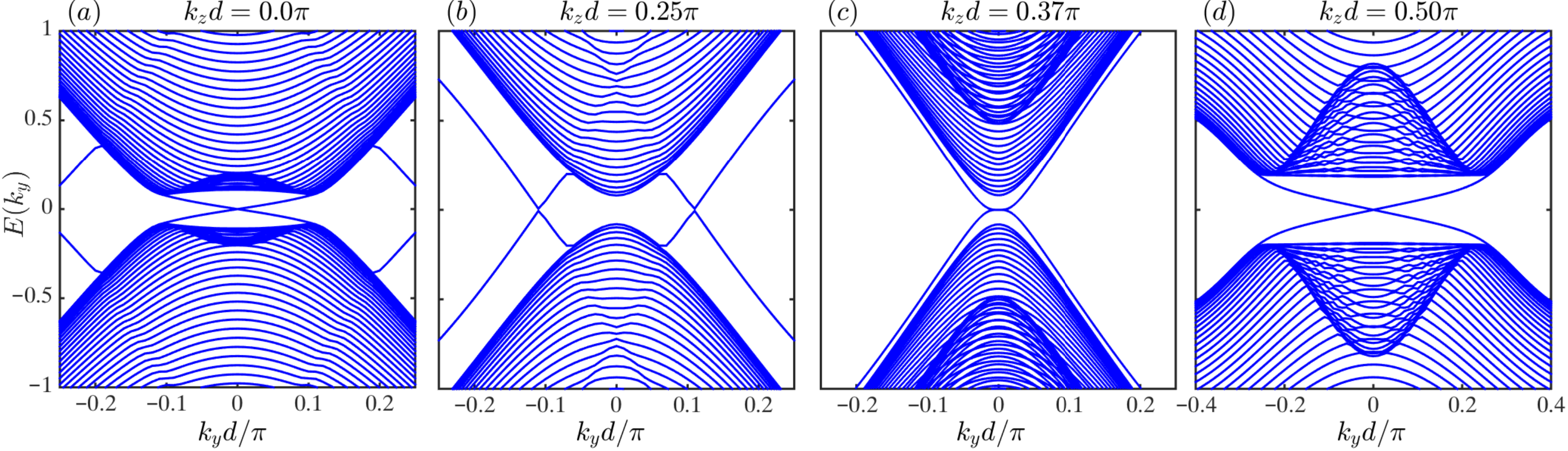}
\caption{(Color online) Band structure of the Weyl
  superconductor as a function of $k_y$ for $\gamma=3t$ (corresponding to Fig. \ref{fig2}(d)) and for different values of $k_z d/\pi \in [0.0, 0.25,
  0.37, 0.5]$. Other parameters are set equal to those of Fig.~\ref{fig2}.}
\label{fig3}
\end{figure*}

\section{numerical simulations}\label{sec:theor}
We now investigate numerically the 
minimal model of a Weyl semimetal with only two 
Weyl points in the band structure, introduced in Eq.~(\ref{eq:Hamkz}). 
The corresponding tight-binding limit of the Hamiltonian $H({\bf k})$ 
can be written as
\begin{eqnarray} \label{eq1}\nonumber
H(\mathbf{k})&=& (2t\sigma_z-\gamma)(\cos Q d-\cos k_zd)+2t\sigma_x\sin k_yd~~~\\
&-&2t \sigma_y\sin k_xd+m\sigma_z (\cos k_yd+\cos
k_xd-2),
\end{eqnarray} 
where $\sigma$ denotes
the  spin degree of freedom on each lattice site, 
$d$ is the lattice constant, 
and $(t, m, \gamma)$ denote hopping energy integrals between two neighboring sites of the lattice. 
This choice of Hamiltonian ensures that the zero-momentum inter-valley superconducting pairing is energetically
more %kh
favorable than the finite momentum intra-valley pairing as discussed in Ref.~\onlinecite{Weyl_SC_ours}. 
In the low-energy limit one can find a 
correspondence between the lattice and continuum model parameters as $\beta = -\gamma \sin Qd$, 
$v_{z} = 2td \sin Qd$, and $v_{\perp} = 2td $. 
We note that positive values of $\gamma$, 
which will be considered further, describe the
tilting of Weyl cones in opposite directions. 
To have  a better understanding of the band structure properties and
phase transitions in the model Hamiltonian used in Eq. (\ref{eq1}),
we diagonalize the Hamiltonian along the $k_z$-axis for a bulk
material. The resulting eigenvalues are given by: 
\begin{eqnarray}\label{dispm}\nonumber
\varepsilon_{\pm,s}(k_z) &=& \pm 2 t(\cos k_zd-\cos Qd)\\
&+& s \sqrt{(\mu-\gamma(\cos k_zd-\cos
  Qd))^2+|\Delta|^2}.~~
\end{eqnarray}
The location of the 
Weyl nodes can be
found by determining the roots of Eq.~(\ref{dispm}), that is,
$\varepsilon(k_z)=0$. The zeros of Eq.~(\ref{dispm}) are
straightforwardly given by 
\begin{eqnarray}\label{sol}
&& k_z^{\pm}= \nonumber \\ 
&&\cos^{-1}
\hspace{-.1cm}
\left\{\frac{(4t^2-\gamma^2)\cos
Qd-\mu\gamma\pm
\hspace{-.1cm}
\sqrt{4\mu^2t^2+(4t^2-\gamma^2)|\Delta|^2}}{4t^2-\gamma^2}\right\}. \nonumber \\ 
\end{eqnarray}
As seen, the absolute value of the
expression in the argument of $\cos^{-1}$ can be larger
than unity, depending on the parameters used. 
Therefore, generally, there arises a region of parameter space in which $\varepsilon(k_z)=0$
has no solution.

To gain insight into the band structure of the Weyl superconductor, we 
choose two plane surfaces of the sample to be aligned parallel to each other. 
The $x$-axis is chosen to be perpendicular to these planes and 
consequently  $k_y$ and $k_z$ are good quantum numbers. 
This permits exact 
diagonalization of the model Hamiltonian, Eq.~(\ref{eq:Hamkz}). 
Without loss of  
generality,  we set the
superconducting gap to $\Delta=0.2t$, and choose a nonzero chemical
potential of $\mu=0.5t$. Thus, we do not consider $\Delta$ as a tilting parameter self-consistently. 
We also define the two Weyl points by setting $Q=0.4\pi/d$ throughout our numerical
calculations.

Figure~\ref{fig2} illustrates
the band structure  $E(k_z)$ of the system  for
different tilting strengths
$\gamma$ and momentum  $k_y=0$. 
The behavior of the band structure in the $k_y$ direction is discussed  later. %kh
As exhibited  in Fig.~\ref{fig2}(a), when $\gamma=0$, there are four point nodes in the superconducting gap, 
with two nodes on each Fermi sheet enclosing two distinct Weyl points.
The point nodes are connected pairwise along the $z$-axis in the momentum-space via the flat band at zero energy, 
corresponding to Majorana surface states. Lines at
$E(k_z)\approx \pm \mu$ describe the
dispersion of the Fermi arc surface states, which connect two distinct Fermi sheets, consistent with Refs.~\onlinecite{Tanaka} and \onlinecite{Weyl_SC_ours}. 

When the tilt parameter $\gamma$ increases within the region $\gamma<2t$, 
the outer gap nodes move towards the corners of first Brillouin
zone, as seen in Fig.~\ref{fig2}(b). At the same time, the Fermi arcs in the particle
and hole sectors move towards each other before eventually crossing. 
%This trend is due to the particle-hole symmetry breaking produced by the tilt.
This crossing corresponds to the switches of particle-hole symmetric and antisymmetric parts of the Fermi-arcs.
From the first term in Eq.~(\ref{eq1}),
it is clear that the flat band reaches the edges of the first Brillouin zone at the phase transition to
a type-II Weyl metal at $\gamma\approx2t$. % kh check 
When $\gamma>2t$ and the system resides in
the type-II Weyl metal phase, we find that the Fermi
arcs hybridize with the bulk bands while the two gap nodes connected by a flat band
reappear from the center of the first Brillouin zone (Figs.~\ref{fig2}(c) and (d)).  %kh
%The findings are interesting from two view points: By
%increasing the nodes' tilting strength $i$) the region in the momentum space where the zero mode Majorna flat
%bands appear increase $ii$) the Fermi arcs are outside of the game. The former
%point suggests a way to increase the region of appearance of Majorana
%fermions from $\approx 0.2\pi$ at $\gamma=0.0$ to $0.6\pi$ at $\gamma=2.5t$. The latter indicates that the
%localized states at the two surfaces due to the Fermi arcs are no
%longer protected by the bulk gap. In other words, the Fermi arcs are now
%hybridized with the bulk bands and do not support protected edge modes. Through
%considering $i$) and $ii$), one can conclude that any edge mode in the
%type-II phase is Majorana mode. By increasing further the tilting
%parameter up to
%$\gamma=3.0t$, the Weyl nodes start to reappear from the center of
%first Brillouin zone. The gap in the band structure at $k_zd=0.0$
%closes at a single point first. By increasing $\gamma$, the gap
%reopens with now a zero mode flat band connecting the appeared Weyl nodes. 
As seen in Fig.~\ref{fig2}(d),
for $\gamma=3.0t$, the Weyl nodes have emerged from the center of
first Brillouin zone, and the
 band structure
has a zero mode flat band that connects the  Weyl nodes.
%Fig.~\ref{fig2}d.  
Focusing on positive values of $k_z$ for concreteness, one observes that the
superconducting gap has two nodes in the bulk at $k_zd/\pi \approx 0.23$ and at $k_zd/\pi \approx
0.37$. A zero energy Majorana flat band exists in two intervals: $k_zd\in [0, 0.23]$ and $k_zd/\pi \in [0.37, 1]$ for $k_y=0$.

\begin{figure*}[t!]
\includegraphics[width=18.0cm,height=10.40cm]{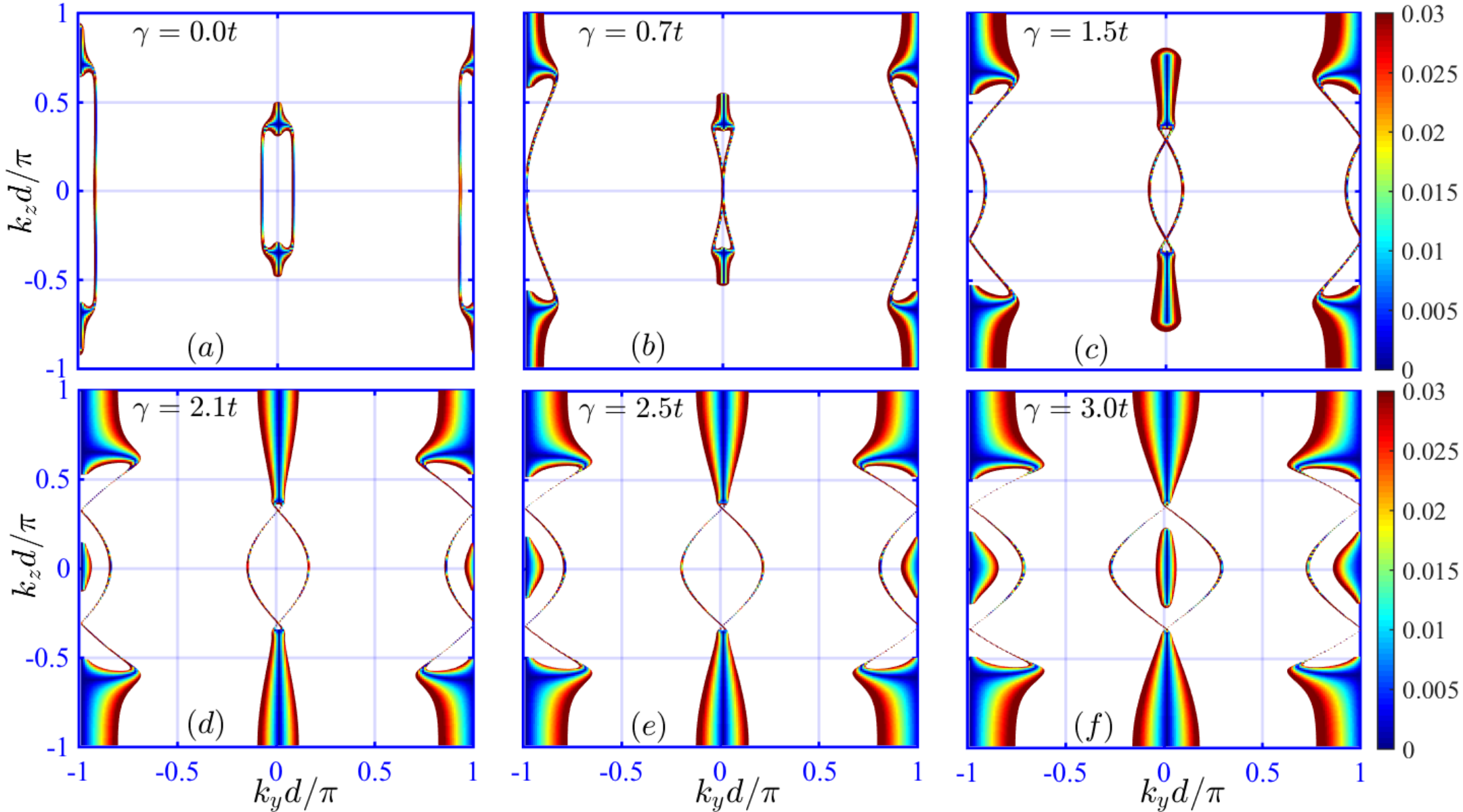}
\caption{(Color online) Two-dimensional momentum space map of the band structure
  at low energies close to zero $E(k_y,k_z)\ll 1$ as a function of
  $k_y$ and $k_z$. The node  titling parameter $\gamma$ evolves from $0$ to
  $3.0t$ corresponding to Figs.~\ref{fig2} and \ref{fig3}. The
  chemical potential and superconducting gap are fixed at
  $\mu=0.5t$ and $\Delta=0.2t$, respectively.}
\label{fig4}
\end{figure*}

In Fig.~\ref{fig3}, we  
present the band structure of a Weyl superconductor again for
$\gamma=3t$, but now as a function of $k_y$ at certain points along the $k_z$-axis. 
The band dispersion along the
$k_y$-axis for $k_z=0$ is shown in Fig.~\ref{fig3}(a).
Here, the bulk spectrum is gapped. 
The two curves that cross zero energy at $k_y=0$ describe the
dispersion of two chiral Majorana states bound to opposite plane surfaces of  the
Weyl superconductor.
At $k_zd/\pi = 0.25$, the
bulk states are gapped as shown in Fig.~\ref{fig3}(b), while there are two Fermi arcs that cross zero energy at $k_yd/\pi \approx 0.1$. 
Near the Weyl point $k_zd/\pi\approx 0.37$,
Fig.~\ref{fig3}(c) shows hints of a flat band in the $k_y$ direction as well. 
However, as seen, the flat band region is very narrow due to the tilted cones. 
The rightmost panel, Fig. \ref{fig3}(d), illustrates that the single point band crossing feature is 
eventually recovered
at $k_zd/\pi=0.5$. 

To provide additional details of the
bulk and surface state dispersion, in  Fig.~\ref{fig4} we present 
a two-dimensional map of the band
structure at energies close to zero. 
In this
figure, the band structure is plotted as a function of $k_y$ and $k_z$
for tilt parameter values ranging from $\gamma=0$ to $3t$. 
When $\gamma=0$, Majorana flat bands exist
inside two Fermi sheets enclosing the Weyl points at $\mathbf{k}=(0,0,\pm Q)$. 
Namely, in the intervals $k_zd/\pi \in [0.35, 0.5]$ and $k_zd/\pi \in [-0.5, -0.35]$, with $k_y=0$. There are also flat bands at
$k_zd/\pi = Qd/\pi\equiv \pm 0.4$, crossing  the $k_z$-axis in the $y$-direction. 
The two Fermi sheets enclosing the Weyl points are 
connected by two Fermi-arcs.

With increases of the tilt parameter $\gamma$, Fermi arcs
become closer to each other and at $\gamma \approx 0.7t$ touch each other at $k_z=0$. 
Increasing $\gamma$ 
further causes the Majorana flat band at $k_y=0$ to extend into the $k_z$ direction,
while its counterpart at $k_z=0$ shrinks in the $k_y$ direction. 
Interestingly, the crossing of the Fermi arcs occurs closer to the gap nodes, %kh
which is maximal above the phase transition at $\gamma\approx 2t$. %kh
At $\gamma\gtrsim2.7t$ one observes a reappearance of gap nodes connected by a Majorana flat band in the middle of first Brillouin zone. 
We note also that a flat band at $k_zd/\pi =\pm Q$ has diminished considerably in the $k_y$ direction.

Thus far we have considered the situation where the tilting parameter is
positive, $\gamma>0$. In the opposite case, 
$\gamma<0$, two gap nodes that are positioned closer to the center of the Brillouin zone, move towards $k_z=0$ as
$|\gamma|$ increases. It is evident that the Fermi arcs hybridize directly with the bulk bands without
any crossing. The two inner cones finally merge into one and open up a gap, such that only a flat band connecting two outer cones remains. 
With further increases in $\gamma$, gap nodes reappear from the edges of the first Brillouin zone. 

We note that the sign of the chemical potential may play
a similar role as $\gamma$, depending on the
doping type that is
incorporated into the system. Therefore, the zero doping level, i.e. $\mu=0$, is a specific point. 
In this regime, the Fermi arcs are
degenerate and constitute a flat band at zero energy connecting the
merged cones at zero energy. Increasing $\gamma$ causes splitting of
the degenerate Fermi arcs until they hybridize with the bulk bands at values
around $\gamma\approx t$. Further increase of $\gamma$ gaps out the
touching gap nodes, and finally at large values of the tilting
parameter e.g. $\gamma=3.75t$ we recover two nodes connected by a flat band.

\section{conclusion}\label{sec:conc}
We now comment on the effect of tilting the Weyl cones in the same direction,
i.e., we consider finite $\alpha$. Note that this tilt breaks inversion symmetry in the system.
Thus, it can be easily shown that  increasing  the tilt parameter $|\alpha|$ suppresses 
the superconducting gap and decreases the critical temperature, 
which is opposite to the effect of tilting the Weyl cones in opposite directions to one another.

The effect of tilting the Weyl cones along the same direction is to split each node in the superconducting gap function into two. 
Taking the limit $\beta \rightarrow 0$ together with 
$|\alpha|>0$ and $|\alpha|\neq 1$ for concreteness, 
situates the four point nodes around each Weyl point given by $\mathbf{k}=(0,0,k_{z,\pm,s})$, where $v_z(k_{z,\pm,s}-Q)=\pm \sqrt{|\Delta|^2+\mu^2}/(1+s\alpha)$ and $s=\pm 1$.  The spectrum of Bogoliubov quasiparticles along the direction of the tilt is given by $ E_{\pm,s} = (\alpha- s)v_z\delta k_z$, bearing a similarity with the type-II Weyl fermions. At the phase transition defined by $|\alpha|=1$, there are two nodes around each Weyl point. These interesting
type-II Bogoliubov quasiparticles could be studied via proximity effect in the superconductor -  type-II Weyl semimetal - superconductor junctions.

Finally, the spectrum of Majorana edge states in the limit $|\alpha|<1$ is given by,
\begin{equation}\label{MFS}
E_{M,\pm}=v_z(k_z-Q) \left [\alpha \pm \frac{|\Delta| v_{\perp} k_y}{\mu\sqrt{\mu^2-v_{\perp}^2 k_y^2}}\right].
\end{equation}
Clearly, the flat band at $k_y=0$ transforms into a linear momentum dispersion.

In conclusion, we have studied superconductivity in a type-II Weyl semimetal. We have found that a quantum phase transition
arises
whereby the superconducting gap nodes disappear by escaping towards  the edges
 or  the center of the first Brillouin zone, and
reappearing at the zone's center, or edges, respectively. The regions of Majorana flat bands might be extended as Fermi
arcs that connect different Fermi sheets enclosing Weyl points hybridize with the bulk bands.
We also note that the superconducting critical temperature and
electronic specific heat might be enhanced at the phase transition towards a type-II Weyl state. 
%Signatures of realization of type-II Weyl nodes   \cite{weyl2sc_expnew}

\section{acknowledgements}
We thank Anton Burkov for the support of this project.
AAZ is financially supported by the Swedish Research Council Grant No. 642-2013-7837 
and by the Goran Gustafsson Foundation. K.H.  is  supported  in part by  ONR  and  a  grant  of  HPC  resources  from  the  DOD
HPCMP. MA is supported by Iran's National Elites Foundation (INEF).

\bibliography{BibWSC}{}

% ==============================================================================
\end{document}